\documentclass[twocolumn,trackchanges]{aastex631}

\usepackage{amsmath}
\usepackage{color}
\hypersetup{
 colorlinks=true,%
 linkcolor=black,
 citecolor=blue,
}

\def\HK#1#2{\textcolor{black}{#2}}
\def\HKK#1#2{\textcolor{black}{#2}}

\begin{document}

\title{Comprehensive Synthesis of Magnetic Tornado: Co-spatial Incidence of Chromospheric Swirls and EUV Brightening}

\author[0000-0003-1134-2770]{Hidetaka Kuniyoshi}
\affiliation{Department of Earth and Planetary Science, The University of Tokyo,
7-3-1 Hongo, Bunkyo-ku, Tokyo, 113-0033, Japan}

\author[0000-0002-2180-1013]{Souvik Bose}
\affiliation{Lockheed Martin Solar \& Astrophysics Laboratory, Palo Alto, CA 94304, USA}
\affiliation{Bay Area Environmental Research Institute, NASA Research Park, Moffett Field, CA 94035, USA}
\affiliation{Institute of Theoretical Astrophysics, University of Oslo, PO Box 1029, Blindern 0315, Oslo, Norway}
\affiliation{Rosseland Centre for Solar Physics, University of Oslo, PO Box 1029, Blindern 0315, Oslo, Norway}

\author[0000-0001-5457-4999]{Takaaki Yokoyama}
\affiliation{Astronomical Observatory, Kyoto University, Sakyo-ku, Kyoto, 606-8502, Japan}



\begin{abstract}

Magnetic tornadoes, characterized as impulsive Alfv{\'e}n waves initiated by photospheric vortices in intergranular lanes, are considered efficient energy channels to the corona. 
Despite their acknowledged importance for solar coronal heating, their observational \HK{proxies}{counterparts} from the corona have not been well understood. 
\HKK{To address this issue, we model an entire coronal loop with footpoints rooted in the \HK{photosphere}{upper convection zone by a radiative MHD simulation}}{ 
To address this issue, we use a radiative MHD simulation of a coronal loop with footpoints rooted in the upper convection zone}
and synthesize the chromospheric and coronal emissions corresponding to a magnetic tornado. Considering SDO/AIA 171 {\AA} and Solar Orbiter/EUI 174 {\AA} channel\HK{}{s}, our synthesis reveals that the coronal response to magnetic tornadoes can be observed as an EUV brightening of which \HK{size}{width} is $\sim 2 \ \mathrm{Mm}$. This brightening is located above the synthesized chromospheric swirl observed in \ion{Ca}{2} 8542 {\AA}, \ion{Ca}{2} K, and \ion{Mg}{2} \HK{K}{k} lines, which can be detected by instruments such as SST/CRISP, GST/FISS, and IRIS.
Considering the height correspondence of the synthesized brightening, magnetic tornadoes can be a mechanism for \HK{the campfires}{the small-scale EUV brightenings such as the solar ``campfires''}. Our findings indicate that coordinated observations encompassing the chromosphere to the corona are indispensable for comprehending the origin of coronal EUV brightenings.

\end{abstract}

\keywords{Solar coronal heating (1989), Solar chromosphere (1479), Radiative magnetohydrodynamics (2009)}


\section{Introduction} \label{sec:intro}

The solar coronal heating problem remains a major question in astrophysics. Why do coronal temperatures, exceeding $1,000,000 \ \mathrm{K}$, rise hundreds of times higher than photospheric temperatures, which are around $\sim~6,000 \ \mathrm{K}$ \citep{Edlen_1943_ZAP}? While previous studies have revealed that the magnetic field plays a dominant role in the heating \citep[e.g.,][]{Parker_1983_ApJ, Pevtsov_2003_ApJ}, the detailed mechanism is still under investigation \citep[see reviews by][]{Klimchuk_2006_SoPh, VanDoorsselaere_2020_SSRv}. 
Of particular interest in this letter is the energy transfer mechanism to the corona, which compensates for the radiative and conductive losses from the corona \citep{Withbroe_1977_ARAA, DiazBaso_2021_AA}.

In the past decades, self-consistent modeling of the energy transfer system has become feasible through the so-called magneto-convection simulations\HK{:}{, i.e.,} radiative magnetohydrodynamic (MHD) simulations, encompassing the upper convection zone, photosphere, chromosphere, and the corona \citep{Leenaarts_2020_LRSP}. 
\noindent
These simulations have self-consistently revealed that not only granular-scale ($\sim 1,000 \ \mathrm{km}$) but also smaller-scale ($< 200 \ \mathrm{km}$) convective motions within intergranular lanes supply a considerable Poynting flux for coronal heating \citep{Rempel_2017_ApJ, Breu_2022_AA}.
Magnetic tornadoes, identified as impulsive Alfv{\'e}n waves originating from photospheric vortex motions within intergranular lanes \citep{Wedemeyer_2012_Nature,Battaglia_2021_AA}, display \HK{proxies}{signatures} observed in the photosphere or chromosphere, manifesting as swirling plasma motions \citep[diameter $\sim 2 \ \mathrm{Mm}$,][]{Shetye_2019_ApJ, Dakanalis_2022_AA}. Magneto-convection simulations focusing on the quiet Sun have indicated that magnetic tornadoes can contribute approximately $50\%$ of the total Poynting flux into the corona \citep{Kuniyoshi_2023_ApJ, Silva_2024_ApJ}.

Unlike the photospheric and chromospheric observations, the coronal response to magnetic tornadoes has not been well understood. 
\citet{Wedemeyer_2012_Nature} have detected EUV brightenings over chromospheric swirls using coordinated observations by Atmospheric Imaging Assembly \citep[AIA;][]{Lemen_2012_SoPh}/Solar Dynamics Observatory \citep[SDO;][]{Pesnell_2012_SoPh} and Swedish 1-m Solar Telescope \citep[SST;][]{Scharmer_2003_SPIE}/CRisp Imaging SpectroPolarimeter \citep[CRISP;][]{Scharmer_2008_ApJ}.
On the other hand, \citet{Tziotziou_2018_AA} have conducted similar coordinated observation and found an EUV darkening above a chromospheric swirl. 
To interpret the coronal response accurately, a comprehensive numerical model capable of addressing both chromospheric and coronal signals of magnetic tornadoes is required. Therefore, in this letter, our objective is to synthesize a magnetic tornado within a coronal loop reproduced in a magneto-convection simulation. 
Unlike our previous simulations, which only considered one half of a loop \citep{Kuniyoshi_2023_ApJ, Kuniyoshi_2024_ApJ}, we now model an entire coronal loop, with the top and bottom boundaries set as the loop footpoints, in accordance with the setup proposed by \citet{Breu_2022_AA}. This approach mitigates numerical wave reflections from the top boundary at the loop apex, thus averting unrealistic modifications to the coronal energy dissipation system.

\section{Methods}

\subsection{Simulation Setup} \label{sec:simulation_setup}

We conduct a three-dimensional magneto-convection simulation using the RAMENS (RAdiation Magnetohydrodynamics Extensive Numerical Solver) code \citep{Iijima_2016_PhD, Iijima_2017_ApJ}. This code solves the compressible magnetohydrodynamic (MHD) equations with gravity, radiation, and thermal conduction. 
The basic equations are given in the conservation form as follows:

\begin{align}
\label{eq:mhd_eqs}
    & \frac{\partial \rho} {\partial t} + \nabla \cdot (\rho \boldsymbol{v} )  = 0, \\
    & \frac{\partial (\rho \boldsymbol{v})}{\partial t} 
      +\nabla \cdot \left[ \rho \boldsymbol{v} \boldsymbol{v}
      + \left( p+\frac{\boldsymbol{B}^2}{8\pi} \right) \boldsymbol{ \underbar I}
      - \frac{\boldsymbol{B} \boldsymbol{B}}{4 \pi}  \right]
      = \rho \boldsymbol{g}, \\
    & \frac{\partial \boldsymbol{B}}{\partial t}+\nabla \cdot (\boldsymbol{vB}-\boldsymbol{Bv})=0, \\
    & \frac{\partial e}{\partial t} 
      + \nabla \cdot \left[
     \left(e+p+\frac{\boldsymbol{B}^2}{8\pi}\right)\boldsymbol{v} -\frac{1}{4\pi} \boldsymbol{B}(\boldsymbol{v}\cdot\boldsymbol{B}) \right] \\
    & =\rho \boldsymbol{g}\cdot \boldsymbol{v}+Q_{\mathrm{cnd}}+Q_{\mathrm{rad}} \nonumber , 
\end{align}

\noindent where $\rho$ is the mass density,  $\boldsymbol{v}$ is the gas velocity, $\boldsymbol{B}$ is the magnetic field, $e=e_{\rm int} + \rho \boldsymbol{v}^2/2 + \boldsymbol{B}^2/8\pi$ is the total energy density, $e_\mathrm{int}$ is the internal energy density,
$p$ is the gas pressure, $\boldsymbol{g}$ is the gravitational acceleration, and $\boldsymbol{\underbar I}$ is unit tensor.
$Q_\mathrm{cnd}$ and $Q_\mathrm{rad}$ denote the heating by thermal conduction and radiation, respectively. $Q_{\mathrm{cnd}}$ is Spitzer-type \HK{}{anisotropic} thermal conduction. 
The radiation $Q_{\rm rad}$ is determined through a combination of optically thick and thin components using a bridging law \citep{Iijima_2016_PhD}. \HK{}{For optically thick part, we solved the radiative transfer equation under the gray local thermodynamic equilibrium (LTE) assumption.} The optically thin part is derived from the radiative loss function retrieved from the CHIANTI atomic database version 7.1 \citep{Dere_1997_AAS, Landi_2012_ApJ}, assuming the coronal abundance. Since the loss function is defined for temperatures $T \geq 10^4 \ \mathrm{K}$, we extrapolate the loss function to the lower temperature range ($\leq 10^4 \ \mathrm{K}$) following the method described in \citet{Goodman_2012_ApJ}.
To close the system, the equation of state is calculated under LTE assumption, considering the six most abundant elements in the solar atmosphere (H, He, C, N, O, Ne). 

We modified the original RAMENS code to accommodate an entire coronal loop without considering the loop curvature, following the methodology outlined in \citet{Breu_2022_AA}. The simulation domain spans a horizontal extent of $6 \ \mathrm{Mm} \times 6 \ \mathrm{Mm}$ in the $xy$-direction, with a vertical extent of $28 \ \mathrm{Mm}$ in the $z$-direction ($-2 \ \mathrm{Mm} \leq z \leq 26 \ \mathrm{Mm}$). 
The top ($z=26 \ \mathrm{Mm}$) and bottom ($z=-2 \ \mathrm{Mm}$) boundaries correspond to the upper convection zone. The upper convection zones have a depth of $2 \ \mathrm{Mm}$ below the optical depth $\tau$ unity located at $z=0 \ \mathrm{Mm}$ and $24 \ \mathrm{Mm}$. 
Following \citet{Breu_2022_AA}, we assume a half-circle loop model for the gravitational acceleration as follows

\begin{align}
    \boldsymbol{g} = -\frac{g \cos{\theta}}{(1 + h/R_{\mathrm{sun}})^2} \hat{\boldsymbol{z}},
\end{align}

\noindent 
where $g=2.74 \times 10^4 \ \mathrm{cm \ s^{-2}}$, $R_{\mathrm{sun}} = 6.96 \times 10^{10} \ \mathrm{cm}$, $\theta = z/r$, $h=r \sin{\theta}$, $r=L_z/\pi$, $L_z = 24 \ \mathrm{Mm}$, and $\hat{\boldsymbol{z}}$ is the unit vector in the $z$-direction.
At the top and bottom boundaries, we impose open boundary conditions for outflows, while fixing the entropy of the inflows through the boundaries to mimic convective energy transport from the deep convection zone \citep[for full details, see][]{Iijima_2016_PhD}. In the $x$- and $y$-directions, periodic boundary conditions are applied.
The grid size is $64 \ \mathrm{km}$ in the $xy$-direction and $60 \ \mathrm{km}$ in the $z$-direction. 

The initial condition in the convection zones is prescribed by Model S \citep{Christensen-Dalsgaard_1996_Science}. Above the surfaces, the initial condition consists of an isothermal stratification permeated by a uniform vertical magnetic field with a strength of $10 \ \mathrm{G}$.
The convection is allowed to relax after $3 \ \mathrm{hr}$ of integration. During this period, a thermal conductive flux from the top boundary is imposed to maintain the coronal temperature above $1 \ \mathrm{MK}$. Subsequently, this conductive flux is removed, and integration continues for another $1.5 \ \mathrm{hr}$ to allow the corona to be heated self-consistently by the Poynting flux from both photospheres. We analyze the last $30 \ \mathrm{min}$ of this period with snapshots taken at intervals of $6 \ \mathrm{s}$. Here, we define $t=0 \ \mathrm{s}$ as the start time of this analysis.


\begin{figure*}[!t]
  \centering
  \includegraphics[width =15cm]{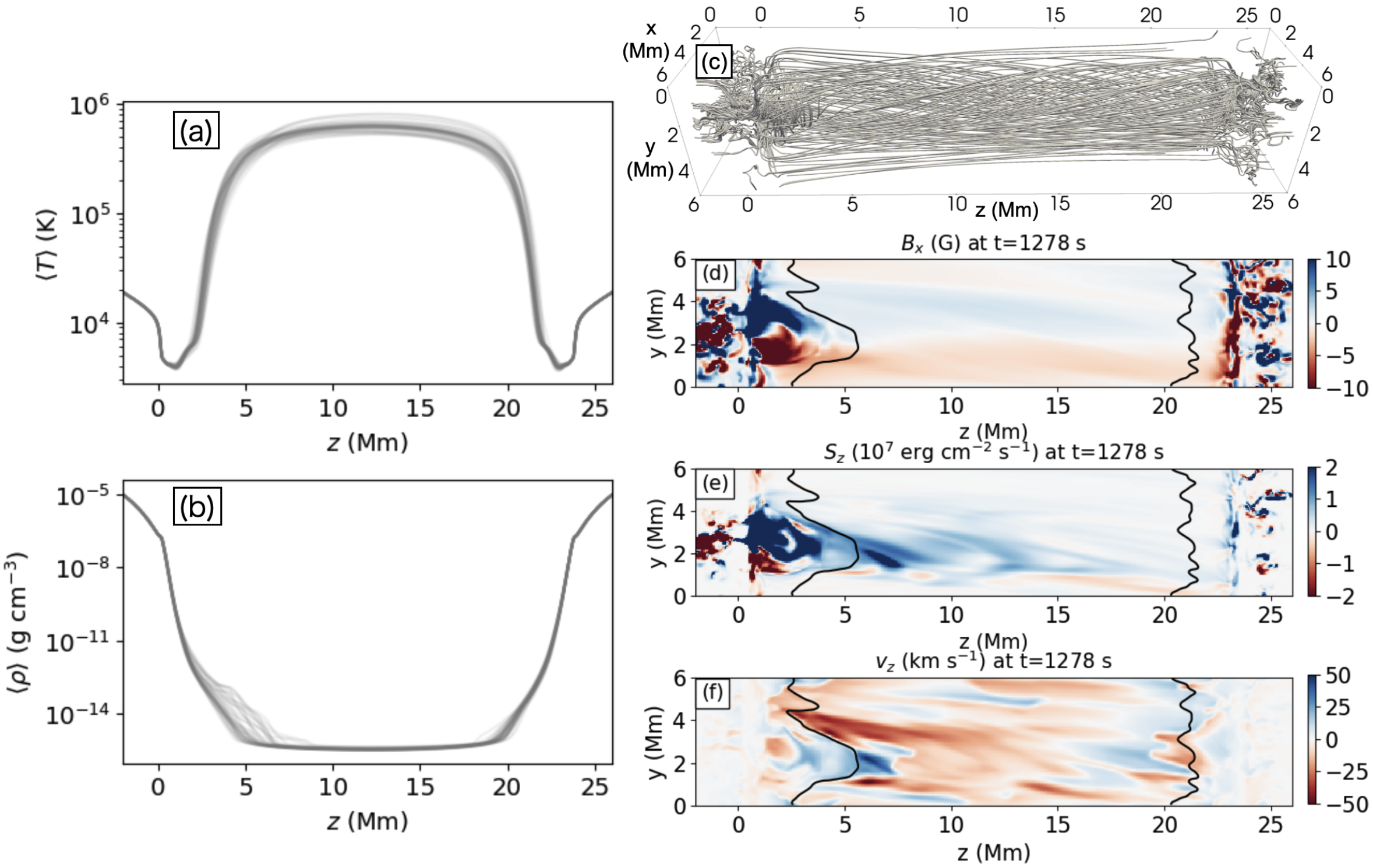}
  \caption{Panel (a) and (b): the probability distributions of the $xy$-averaged temperature $\langle T \rangle$ and mass density $\langle \rho \rangle$ for $t$ and $z$. Panel (c)--(f): snapshots (\HK{}{$t=1,278 \ \mathrm{s}$}) of magnetic field lines, $B_y$, $S_z$, and $v_z$ on the $xz$-plane at $x=4.2 \ \mathrm{Mm}$ in the presence of the magnetic tornado. The black contours in panel (d)--(f) indicate the transition region height where $T=80,000 \ \mathrm{K}$.}
  \label{fig:overview}
\end{figure*}

\subsection{Synthetic Emission} \label{sec:synthesis}

Chromospheric swirls have frequently been observed in \ion{Ca}{2} IR and H$\alpha$ by ground-based telescopes such as SST/CRISP \citep[e.g.,][]{Shetye_2019_ApJ, Dakanalis_2022_AA}. Additionally, an observation through the Interface Region Imaging Spectrograph \citep[IRIS,][]{DePontieu_2014_SoPh} has detected swirls in \ion{Mg}{2} line \citep{Park_2016_AA}. In this paper, we synthesize \ion{Ca}{2} 8542 {\AA}, \ion{Ca}{2} K, and \ion{Mg}{2} \HK{K}{k} spectral lines, utilizing the publicly available RH1.5D\footnote{\url{https://github.com/ITA-Solar/rh}} code \citep{2001ApJ...557..389U,Pereira_2015_AA}. \HKK{}{This code can treat optically thick line formation under non-LTE conditions and partial frequency redistribution, which is critical to modelling the chromospheric spectral lines in detail. The 1.5D (column-by-column) treatment of radiation transport is generally valid except at the cores of strong chromospheric lines such as \ion{Ca}{2} K and \ion{Mg}{2} \HK{K}{k} where the effects of lateral radiation (3D) transport become important \citep{2017A&A...597A..46S,2018A&A...611A..62B}. However, since the aim of this paper is not a direct comparison of the synthesized observables with actual observations and owing to the substantially high time complexity of 3D non-LTE radiative transfer, the benefits of the 1.5D approach far outweigh its limitations \citep{Pereira_2015_AA}. Moreover, as shown in Figure \ref{fig:tornado_morphology}, the 1.5D approach distinctly reproduces the signature of the swirls in the chromosphere.}

For the coronal response, we synthesize the EUV emission as observed in the AIA 171 {\AA} channel. Furthermore, the 174 {\AA} channel of the Extreme Ultraviolet Imager \citep[EUI;][]{Rochus_2020_AA} on board Solar Orbiter \citep[SolO\HK{}{;}][]{Muller_2020_AA} is also calculated. 
The coronal emission is calculated assuming the optically thin approximation under ionization equilibrium, following a methodology similar to that of \citet{Chen_2021_AA}. The emission $I_{\mathrm{corona}}$ is given as: 

\begin{align}
\label{eq:thin_emission}
    I_{\mathrm{corona}} = \int n_{\HK{}{\mathrm{e}}} n_{\HK{}{\mathrm{H}}} K(T) ds,
\end{align}

\noindent where $s$ is the line of sight direction, $n_e$ is electron number density, $n_H$ is hydrogen number density, and $K(T)$ is the contribution function corresponding to the AIA 171 {\AA} and EUI 174 {\AA} channels computed using the \HK{CHIANTI database \citep{Dere_1997_AAS, Landi_2012_ApJ}}{FoMo code \citep{VanDoorsselaere_2016_FrASS}}. For a direct comparison between the synthesized emission and observations, we consider the pixel sizes of the instruments. The AIA instrument has a spatial resolution of $\approx$ $1.2\HK{}{"}$, while the EUI instrument has $\approx$ $0.4\HK{}{"}$. Following the procedure outlined by \citet{Breu_2022_AA}, we resample the synthesized emissions by summing up neighboring pixel patches from the numerical model to match the instrumental spatial scale. For simplicity, we did not convolve with the point spread function.

\section{Results}
\label{sec:results}

\subsection{Simulation Overview}

Figure~\ref{fig:overview}a and b depict the probability distribution of horizontally averaged temperature $T$ and mass density $\rho$, denoted as $\langle T \rangle$ and $\langle \rho \rangle$, respectively. \HK{}{It is worth noting that we define the angle brackets $\langle \rangle$ as representing the $xy$-averaging.} $\langle T \rangle$ ranges from $0.5 \ \mathrm{MK}$ to $0.8 \ \mathrm{MK}$, while locally, it exceeds $1 \ \mathrm{MK}$. Furthermore, $\langle \rho \rangle$ reveals that chromospheric plasma on the left-hand side extends to higher altitudes ($z=3$--$6 \ \mathrm{Mm}$) compared to the right-hand side. This result arises from a chromospheric jet initiated, which is revisited later. 

\begin{figure}[!t]
  \centering
  \includegraphics[width =8.5cm]{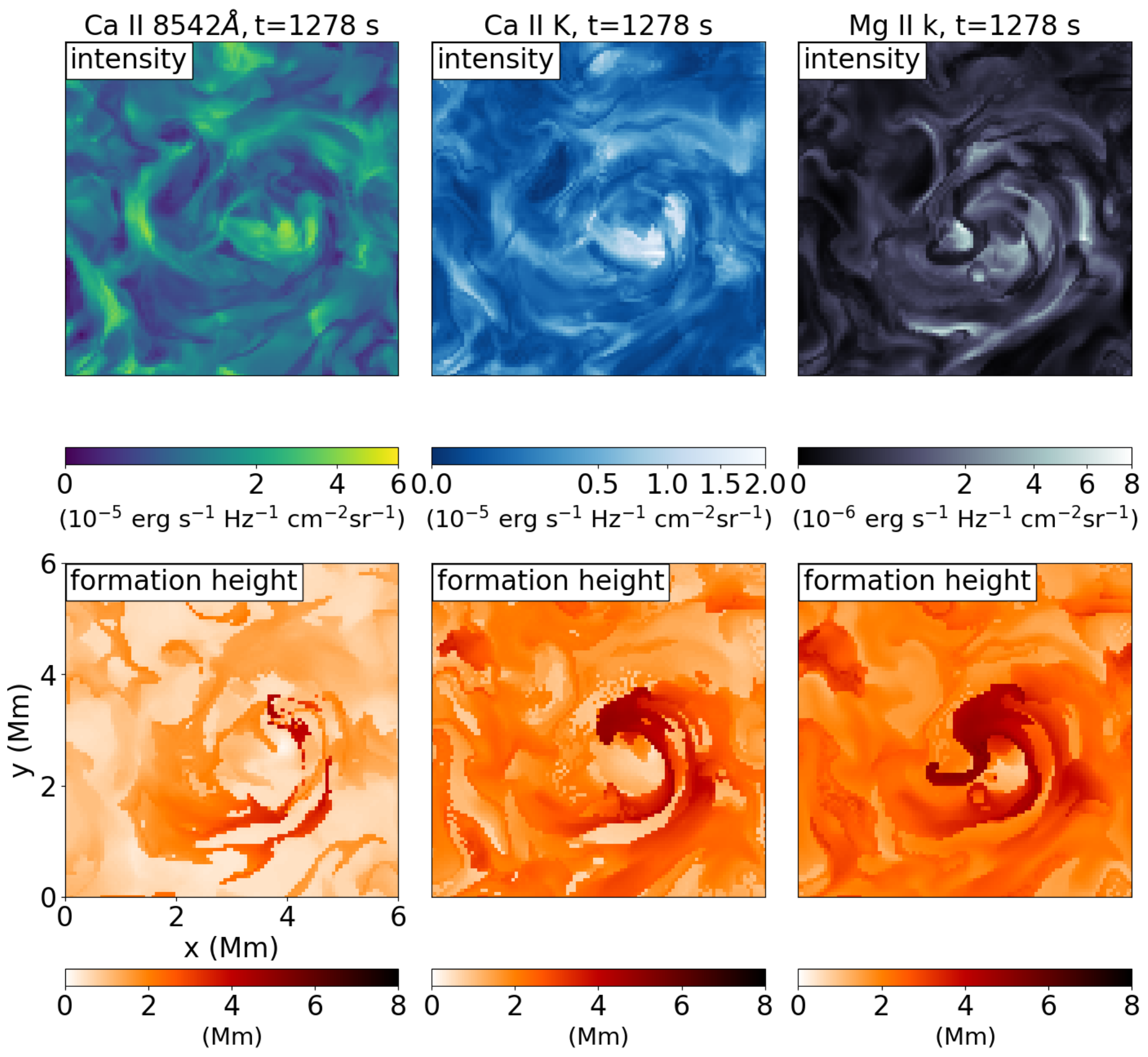}
  \caption{Top row: Snapshots (\HK{}{$t=1,278 \ \mathrm{s}$}) of chromospheric emissions in \ion{Ca}{2} 8542 {\AA}, \ion{Ca}{2} K, and \ion{Mg}{2} \HK{K}{k} as observed from the coronal apex ($z=12 \ \mathrm{Mm}$). Bottom row: The formation heights of the emissions.
  \HK{}{The associated animation shows the temporal evolution over a period from $t=1,254 \ \mathrm{s}$ to $t=1,350 \ \mathrm{s}$.}
  }
  \label{fig:tornado_morphology}
\end{figure}

\HK{}{During the analyzed period, a magnetic tornado is generated by photospheric vortex flows, which is the same triggering mechanism as presented in many previous simulations \citep{Kuniyoshi_2023_ApJ, Silva_2024_ApJ}.} 
The lifetime of the magnetic tornado is approximately $10 \ \mathrm{min} \ (950 \ \mathrm{s} \lesssim t \lesssim 1,550 \ \mathrm{s})$, corresponding to that of the photospheric vortex, which persists for about one granular turnover time.
Figure \ref{fig:overview}c displays a snapshot of magnetic field lines when the magnetic tornado is produced. It exhibits a coherently twisted feature extending from the left-hand side of the surface ($z=0 \ \mathrm{Mm}$) to the entire coronal volume. Figure \ref{fig:tornado_morphology}d--f illustrates snapshots of the $x$-component of the magnetic field $B_x$, vertical Poynting flux $S_z$ (where $S_z=(B_x^2 + B_y^2) v_z / 4 \pi - (v_x B_x + v_y B_y) B_z / 4 \pi$), and vertical velocity $v_z$ on the $yz$-plane at $x=4.2 \ \mathrm{Mm}$, with the transition region height where $T=\HK{40,000}{80,000} \ \mathrm{K}$. These quantities clearly depict typical features of magnetic tornadoes. In the region $1 \ \mathrm{Mm} < z < 3 \ \mathrm{Mm}$, highly twisted $B_x$ with an absolute value reaching \HK{}{over} $10 \ \mathrm{G}$ is observed, extending to the coronal height. Through this region, an amplified vertical Poynting flux $S_z$ exceeding $1 \times 10^7 \ \mathrm{erg \ cm^{-2} \ s^{-1}}$ is channeled into the corona. Additionally, the transition region height increases just above this region to $5 \ \mathrm{Mm}$, from which enhanced vertical velocity $v_z$ surpassing $50 \ \mathrm{km \ s^{-1}}$ extends into the corona. 
They are the features of the chromospheric jet driven by the Lorentz force accompanying the magnetic tornado, which is consistent with the previous simulations \citep{Iijima_2017_ApJ, Dey_2024_arXiv}.

\subsection{Magnetic Tornado Synthesis}

\begin{figure}[!t]
  \centering
  \includegraphics[width =8.5cm]{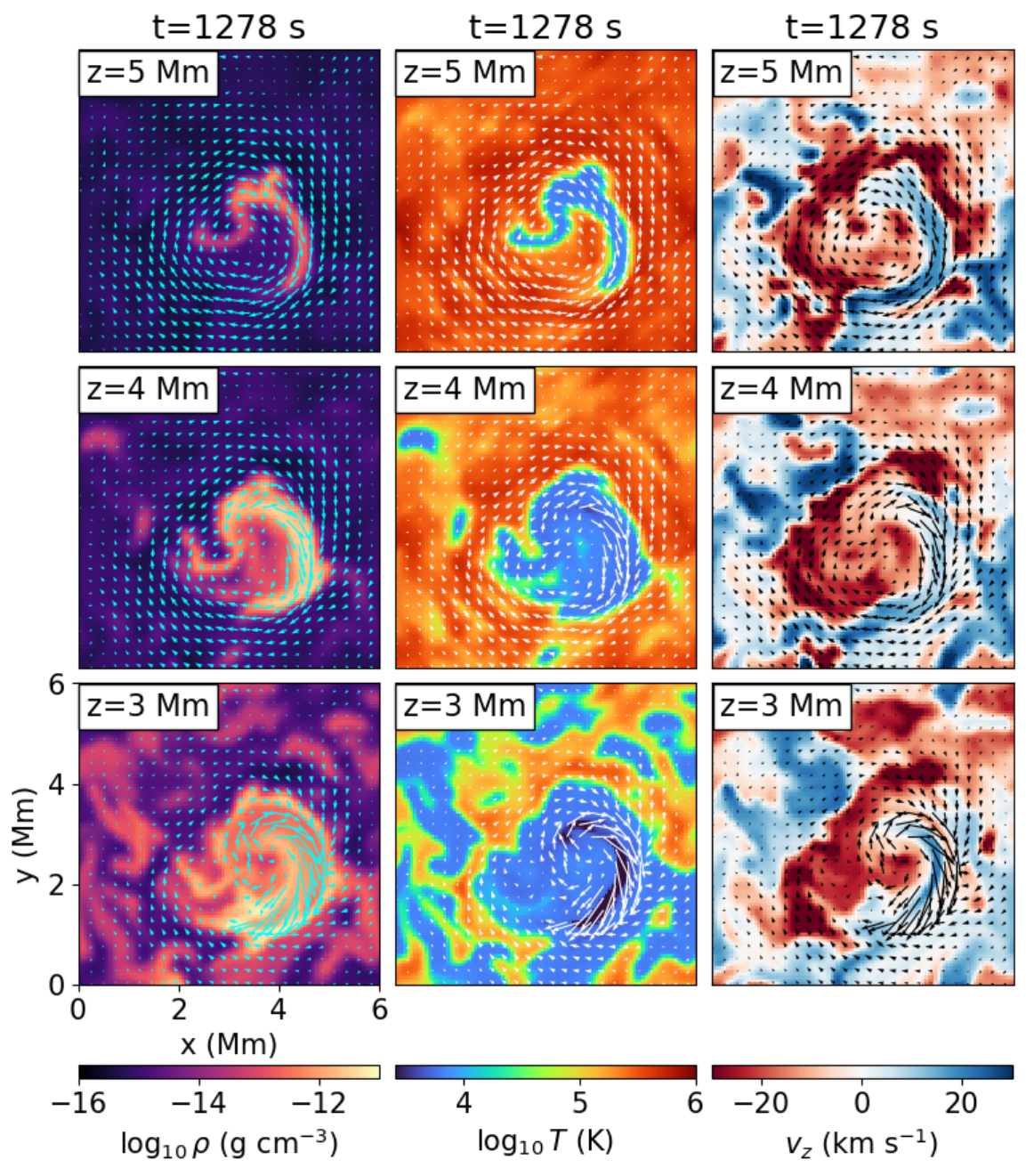}
  \caption{Horizontal slices of $\rho$, $T$, and $v_z$ in color with $(B_x,B_y)$ in arrows at $z=5 \ \mathrm{Mm}$ (top row), $z=4 \ \mathrm{Mm}$ (middle row), and $z=3 \ \mathrm{Mm}$ (bottom row) at \HK{}{$t=1,278 \ \mathrm{s}$.} \HK{}{The associated animation shows the temporal evolution over a period from $t=1,254 \ \mathrm{s}$ to $t=1,350 \ \mathrm{s}$.}}
  \label{fig:tornado_chromosphere}
\end{figure}

In this section, we synthesize the chromospheric and coronal emissions corresponding to the magnetic tornado. 
The first row of Figure \ref{fig:tornado_morphology} displays the chromospheric \HK{emissions}{synthesis} of \ion{Ca}{2} 8542 {\AA}, \ion{Ca}{2} K, and \ion{Mg}{2} \HK{}{k} line cores as observed \HK{along the line of sight from the coronal apex at $z=12 \ \mathrm{Mm}$ to $z=0 \ \mathrm{Mm}$}{from the negative $z$-direction at $z=12 \ \mathrm{Mm}$}.
\HKK{}{These panels and associated animation illustrate that the magnetic tornado can be observed as a \HKK{}{rotating} swirl-like feature with arc structures, with a diameter of about $3 \ \mathrm{Mm}$.}
The second row shows the $\tau=1$ formation height corresponding to the wavelength of the images shown in the top row. \HKK{}{They reveal that the synthesized swirl originating from the magnetic tornado is formed between $z=3$--$5 \ \mathrm{Mm}$.}
\HK{}{In Figure \ref{fig:tornado_chromosphere}}, horizontal slices of $\rho$, $T$, and $v_z$ at corresponding heights $z=3$--$\HK{}{5} \ \mathrm{Mm}$ are displayed, with horizontal magnetic field ($B_x,B_y$) in arrows. These quantities have displayed that the synthesized swirl is characterized by the chromospheric jet with denser and cooler structures than surrounding plasma, upflowing through the twisted magnetic field \HK{}{(see also the corresponding animation).}

\begin{figure}[!t]
  \centering
  \includegraphics[width =8.5cm]{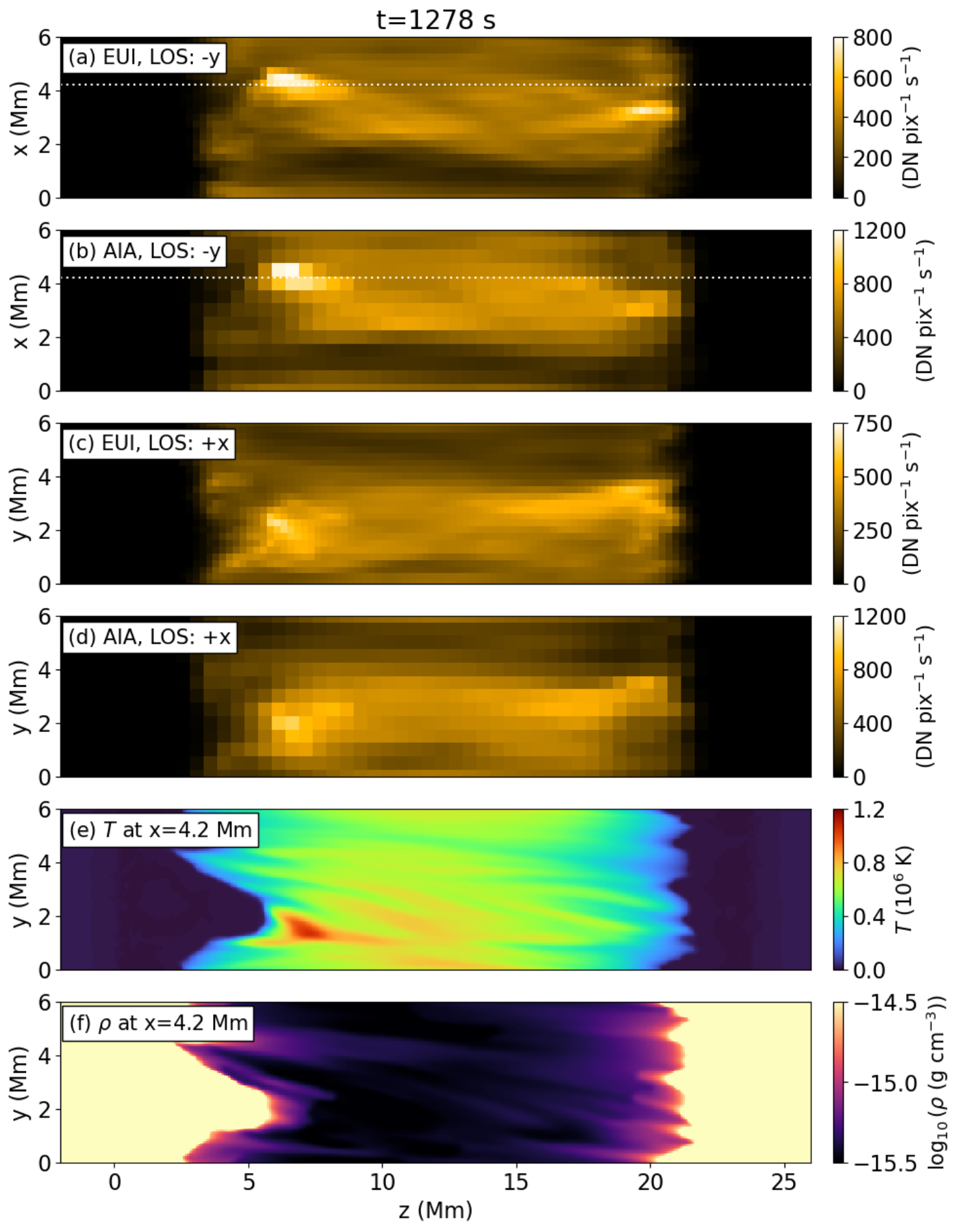}
  \caption{Panel (a)--(d): snapshots (\HK{}{$t=1,278 \ \mathrm{s}$}) of synthesized emissions in the SolO/EUI $174$ {\AA} and SDO/AIA $171$ {\AA} channel as seen from the negative $y$-direction (panel (a) and (b)) and positive $x$-direction (panel (c) and (d).
  Panel (e)--(f): vertical slices of $T$ and $\rho$ through the EUV brightening depicted in panels (a) and (b) ($t=1,278 \ \mathrm{s}$). These slices are taken at $x=4.2 \ \mathrm{Mm}$, indicated by the white dashed lines in panels (a) and (b).
  \HK{}{The corresponding animation shows the temporal evolution over a period from $t=1,254 \ \mathrm{s}$ to $t=1,350 \ \mathrm{s}$.}}
  \label{fig:tornado_corona}
\end{figure}


\HK{}{Figure~\ref{fig:tornado_corona}a--d present the synthesized coronal emission as observed by the SolO/EUI 174 {\AA} and SDO/AIA 171 {\AA} channels from the negative $y$-direction and positive $x$-direction at the same time as the chromospheric lines.
These panels \HK{}{and associated animations} reveal that the EUV brightening occurs within the range of $6 \ \mathrm{Mm} \leq z \leq 9 \ \mathrm{Mm}$ with a width of approximately $1 \ \mathrm{Mm}$, during the period $1,278 \ \mathrm{s} \leq t \leq 1,350 \ \mathrm{s}$}.
Although this simulation disregards the curvature of the coronal loop, assuming the coronal loop is semicircular, the corresponding formation altitude from the surface ranges from $5 \ \mathrm{Mm}$ to $8 \ \mathrm{Mm}$. 
\HKK{}{It is important to note that we do not account for EUV absorption by chromospheric materials (such as \ion{H}{1}, \ion{He}{1}, and \ion{He}{2}), as there are no chromospheric jets or prominences generated in front of the coronal brightening along the line of sight of the synthesized emissions.}

To analyze the thermal properties of this brightening,
Figure \ref{fig:tornado_corona}e--f present the vertical slices of $T$ and $\rho$ at $x=4.2 \ \mathrm{Mm}$, where the center of the brightening locates. These maps and the corresponding animation show that the brightening is caused by the local enhancement of mass and temperature, resulting from the following processes. First, the transition region height around $y=1.5$--$3 \ \mathrm{Mm}$ is lifted due to the chromospheric jet driven by the magnetic tornado (see also Figure \ref{fig:tornado_chromosphere}). This jet then supplies mass to the corona above, which is subsequently heated to over $1 \ \mathrm{MK}$. From these results, we can infer that the coronal response to magnetic tornadoes can be observed as local brightenings above chromospheric swirls.

\begin{figure}[!t]
  \centering
  \includegraphics[width =8.5cm]{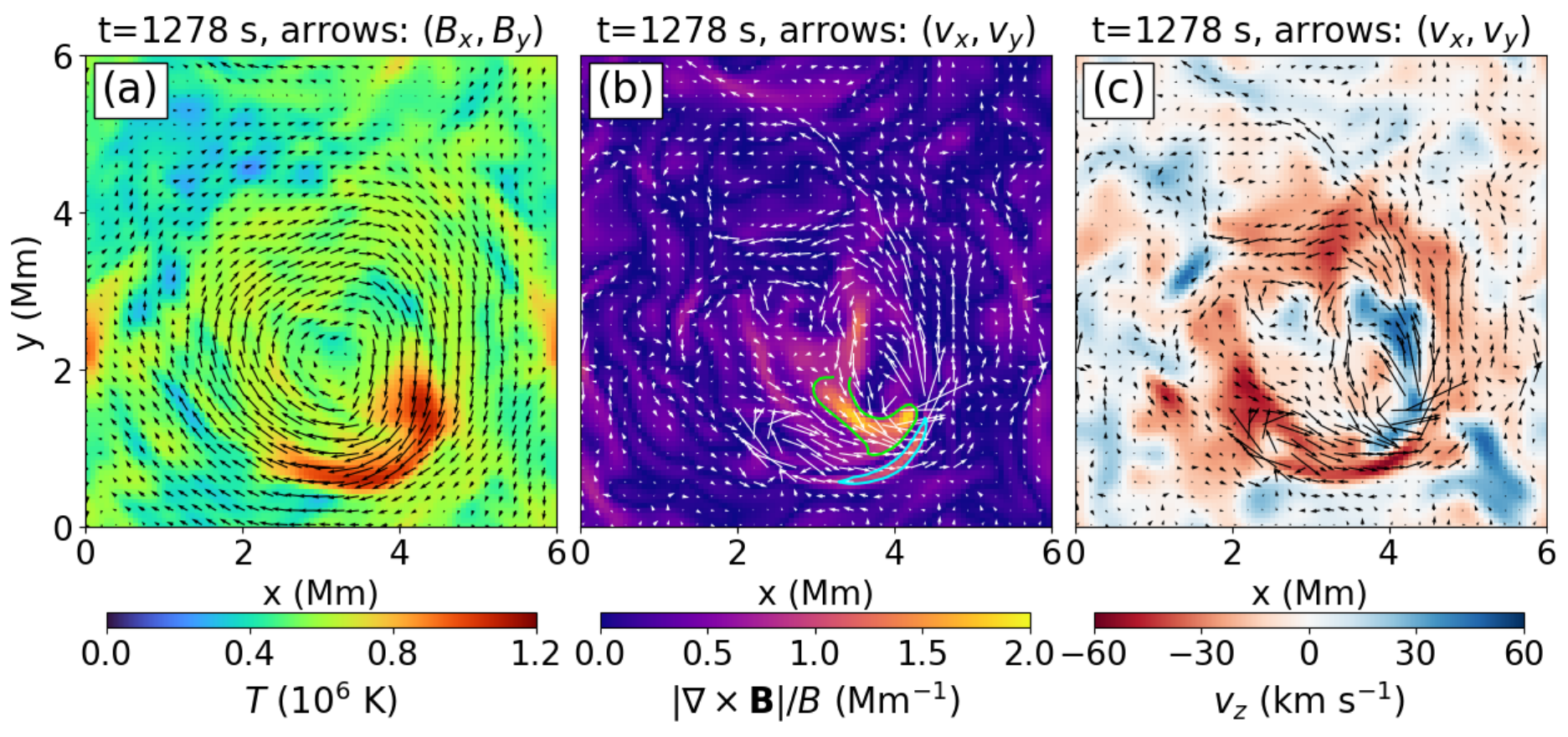}
  \caption{The horizontal slices ($z=7.3 \ \mathrm{Mm}$) of temperature $T$, $|\nabla \times \boldsymbol{B}|/B$ and vertical velocity $v_z$ in color, with horizontal magnetic field $(B_x,B_y)$ or horizontal velocity field $(v_x,v_y)$ in arrows at $t=1,278 \ \mathrm{s}$. }
  \label{fig:eui_aia}
\end{figure}

Figure \ref{fig:eui_aia}a--c display horizontal slices of $T$, $\alpha=|\nabla \times \boldsymbol{B}|/B$, and $v_z$ at $z=7.3 \ \mathrm{Mm}$, where the temperature peak locates at $t=1,278 \ \mathrm{s}$ (see Figure~\ref{fig:tornado_corona}e). Additionally, we plot $(B_x,B_y)$ in panel (a) and $(v_x,v_y)$ in panels (b) and (c). 
The temperature distribution shows that the heating occurs within the twisted magnetic field created by the magnetic tornado. 
We plot $\alpha$ to confirm whether this heating mechanism is due to magnetic reconnection. $\alpha$ highlights potential reconnection regions, as it is maximized where strong current density coincides with weak magnetic field strength.
Around the heated region ($T>1 \ \mathrm{MK}$), two elongated structures with enhanced $alpha$ ($>0.8 \ \mathrm{Mm^{-1}}$) corresponding to current sheets are formed, outlined by cyan (current sheet A) and green lines (current sheet B). 
Furthermore, the velocity patterns around the current sheets indicate additional signatures of magnetic reconnection. We observe inflow into and bi-directional outflow from the current sheet A. Additionally, an outflow-like velocity pattern in the positive $x$-direction emerges from current sheet B. However, the other part of the outflow is not visible as it overlaps with the inflow into current sheet A. One part of the outflow from the current sheet A propagates upward ($v_z>0$) through the background magnetic field, with another part propagating downward ($v_z < 0$). This upflowing heated plasma elucidates the behavior of the EUV brightening, which propagates in the positive $z$-direction (see the animation associated with Figure \ref{fig:tornado_corona}). These properties suggest that the heating mechanism is the component magnetic reconnection between internally adjacent field lines within the twisted magnetic field.

\section{Discussion} \label{sec:discussion}

Following previous studies \citep{Wedemeyer_2012_Nature, Tziotziou_2018_AA}, our synthesis has illustrated that a magnetic tornado can be observed as a swirl in chromospheric lines. 
The synthesized swirl in \ion{Ca}{2} lines corresponds to previous observations by SST/CRISP, exhibiting a similar diameter of $\sim 2 \ \mathrm{Mm}$ \citep[e.g.,][]{Shetye_2019_ApJ}.
Our study suggests that chromospheric swirls can also be observed in \ion{Mg}{2}~k line. This result aligns with previous IRIS observations by \citet{Park_2016_AA}, which revealed the signature of the chromospheric swirl obtained in a sit-and-stare mode of the \ion{Mg}{2} \HK{K}{k} 2796 {\AA} line. Considering that the diameter of our swirl is sufficiently larger than the spatial resolution of IRIS \citep[$0.4\HK{}{"}$][]{DePontieu_2014_SoPh}, our synthesis indicates that swirls can be observed in the slit-jaw images (SJIs) in the \ion{Mg}{2} \HK{K}{k} filter. 

We have also analyzed the coronal response to the magnetic tornado, as observed in the SolO/EUI 174 {\AA} and SDO/AIA 171 {\AA} channels. From our synthesis, a brightening in both channels is observed above the chromospheric swirl. This result well corresponds to the previous observation by \citet{Wedemeyer_2012_Nature}, conducting the coordinated observation by SST/CRISP \ion{Ca}{2} 8542 {\AA} line and SDO/AIA 171 {\AA} channel. On the other hand, our result contradicts the observations by \citet{Tziotziou_2018_AA}, which shows a darkening in AIA 171 {\AA} channel above a chromospheric swirl observed in SST/CRISP \ion{Ca}{2} 8542 {\AA} and H$\alpha$ lines. 
\HKK{}{This discrepancy may arise because chromospheric materials such as coronal rains or fibrils positioned over the magnetic tornado, obscure its coronal brightening.}

The altitude of the synthesized coronal brightening from the surface ranges from $5 \ \mathrm{Mm}$ to $8 \ \mathrm{Mm}$ when we assume our coronal loop to be semi-circular. Interestingly, the lower limit is consistent with the reported formation height of the campfires, at least a portion of which is caused by the coronal EUV brightenings observed by SolO/EUI 174 {\AA} channel \citep{Berghmans_2021_AA}.
Our results suggest that magnetic tornadoes are likely mechanisms for producing campfires. 
This aligns with the previous study suggesting that twisted magnetic field structures may play a crucial role in creating at least a portion of campfires \citep{Chen_2021_AA}.
It should also be noted that the reconnection of low-lying or emerging loops, as well as magnetic flux cancellation, are other promising mechanisms \citep{Berghmans_2021_AA, Panesar_2021_ApJ}.
Recent observations propose that the physics behind all the campfires are likely not the same because some have an IRIS counterpart while others do not \citep{Nelson_2023_AA}. 
Therefore, to elucidate the origin of campfires, coordinated observations by ground-based telescopes such as the SST/CRISP or Goode Solar Telescope \citep[GST;][]{Cao_2010_AN}/Fast Imaging Solar Spectrograph \citep[FISS][]{Chae_2013_SoPh} for the chromosphere and the SolO/EUI for the corona are crucial. If magnetic tornadoes indeed trigger campfires, chromospheric swirls should be observable just below the EUV brightenings.



It is worth noting that spectropolarimetric synthesis of magnetic tornadoes is also performed using the dataset computed by the RAMENS code \citep{Matsumoto_2023_MNRAS}. These studies have predicted arc-like linear polarisation signals originating from the highly twisted magnetic field lines, \HK{}{which can be observed by the upcoming polarimetric observations} such as SUNRISE III/Sunrise Chromospheric Infrared spectro-Polarimeter \citep[SCIP;][]{Katsukawa_2020_SPIE}, Daniel K. Inouye Solar Telescope \citep[DKIST;][]{Rimmele_2020_SoPh} and European Solar Telescope \citep[EST;][]{Quintero_2022_AA}.




\

Numerical computations were carried out on the Cray XC50 at the Center for Computational Astrophysics (CfCA), National Astronomical Observatory of Japan.
\HK{}{S.B. gratefully acknowledges support from NASA grant 80NSSC20K1272 “Flux emergence and the structure, dynamics, and energetics of the solar atmosphere” and from NASA contract NNG09FA40C (IRIS). The optically thick synthesis was performed on resources provided by Sigma2 – the National Infrastructure for High Performance Computing and Data Storage in Norway.}
 T.Y. is supported by the JSPS KAKENHI Grant Number JP21H01124, JP20KK0072, and JP21H04492.
This work was supported by NAOJ Research Coordination Committee, NINS, Grant Number NAOJ-RCC-2301-0301.


\begin{thebibliography}{}
\expandafter\ifx\csname natexlab\endcsname\relax\def\natexlab#1{#1}\fi
\providecommand{\url}[1]{\href{#1}{#1}}
\providecommand{\dodoi}[1]{doi:~\href{http://doi.org/#1}{\nolinkurl{#1}}}
\providecommand{\doeprint}[1]{\href{http://ascl.net/#1}{\nolinkurl{http://ascl.net/#1}}}
\providecommand{\doarXiv}[1]{\href{https://arxiv.org/abs/#1}{\nolinkurl{https://arxiv.org/abs/#1}}}

\bibitem[{{Battaglia} {et~al.}(2021){Battaglia}, {Canivete Cuissa}, {Calvo}, {Bossart}, \& {Steiner}}]{Battaglia_2021_AA}
{Battaglia}, A.~F., {Canivete Cuissa}, J.~R., {Calvo}, F., {Bossart}, A.~A., \& {Steiner}, O. 2021, \aap, 649, A121, \dodoi{10.1051/0004-6361/202040110}

\bibitem[{{Berghmans} {et~al.}(2021){Berghmans}, {Auch{\`e}re}, {Long}, {Soubri{\'e}}, {Mierla}, {Zhukov}, {Sch{\"u}hle}, {Antolin}, {Harra}, {Parenti}, {Podladchikova}, {Aznar Cuadrado}, {Buchlin}, {Dolla}, {Verbeeck}, {Gissot}, {Teriaca}, {Haberreiter}, {Katsiyannis}, {Rodriguez}, {Kraaikamp}, {Smith}, {Stegen}, {Rochus}, {Halain}, {Jacques}, {Thompson}, \& {Inhester}}]{Berghmans_2021_AA}
{Berghmans}, D., {Auch{\`e}re}, F., {Long}, D.~M., {et~al.} 2021, \aap, 656, L4, \dodoi{10.1051/0004-6361/202140380}

\bibitem[{{Bj{\o}rgen} {et~al.}(2018){Bj{\o}rgen}, {Sukhorukov}, {Leenaarts}, {Carlsson}, {de la Cruz Rodr{\'\i}guez}, {Scharmer}, \& {Hansteen}}]{2018A&A...611A..62B}
{Bj{\o}rgen}, J.~P., {Sukhorukov}, A.~V., {Leenaarts}, J., {et~al.} 2018, \aap, 611, A62, \dodoi{10.1051/0004-6361/201731926}

\bibitem[{{Breu} {et~al.}(2022){Breu}, {Peter}, {Cameron}, {Solanki}, {Przybylski}, {Rempel}, \& {Chitta}}]{Breu_2022_AA}
{Breu}, C., {Peter}, H., {Cameron}, R., {et~al.} 2022, \aap, 658, A45, \dodoi{10.1051/0004-6361/20214145110.48550/arXiv.2112.11549}

\bibitem[{{Cao} {et~al.}(2010){Cao}, {Gorceix}, {Coulter}, {Ahn}, {Rimmele}, \& {Goode}}]{Cao_2010_AN}
{Cao}, W., {Gorceix}, N., {Coulter}, R., {et~al.} 2010, Astronomische Nachrichten, 331, 636, \dodoi{10.1002/asna.201011390}

\bibitem[{{Chae} {et~al.}(2013){Chae}, {Park}, {Ahn}, {Yang}, {Park}, {Nah}, {Jang}, {Cho}, {Cao}, \& {Goode}}]{Chae_2013_SoPh}
{Chae}, J., {Park}, H.-M., {Ahn}, K., {et~al.} 2013, \solphys, 288, 1, \dodoi{10.1007/s11207-012-0147-x}

\bibitem[{{Chen} {et~al.}(2021){Chen}, {Przybylski}, {Peter}, {Tian}, {Auch{\`e}re}, \& {Berghmans}}]{Chen_2021_AA}
{Chen}, Y., {Przybylski}, D., {Peter}, H., {et~al.} 2021, \aap, 656, L7, \dodoi{10.1051/0004-6361/202140638}

\bibitem[{{Christensen-Dalsgaard} {et~al.}(1996){Christensen-Dalsgaard}, {Dappen}, {Ajukov}, {Anderson}, {Antia}, {Basu}, {Baturin}, {Berthomieu}, {Chaboyer}, {Chitre}, {Cox}, {Demarque}, {Donatowicz}, {Dziembowski}, {Gabriel}, {Gough}, {Guenther}, {Guzik}, {Harvey}, {Hill}, {Houdek}, {Iglesias}, {Kosovichev}, {Leibacher}, {Morel}, {Proffitt}, {Provost}, {Reiter}, {Rhodes}, {Rogers}, {Roxburgh}, {Thompson}, \& {Ulrich}}]{Christensen-Dalsgaard_1996_Science}
{Christensen-Dalsgaard}, J., {Dappen}, W., {Ajukov}, S.~V., {et~al.} 1996, Science, 272, 1286, \dodoi{10.1126/science.272.5266.1286}

\bibitem[{{Dakanalis} {et~al.}(2022){Dakanalis}, {Tsiropoula}, {Tziotziou}, \& {Kontogiannis}}]{Dakanalis_2022_AA}
{Dakanalis}, I., {Tsiropoula}, G., {Tziotziou}, K., \& {Kontogiannis}, I. 2022, \aap, 663, A94, \dodoi{10.1051/0004-6361/202243236}

\bibitem[{{De Pontieu} {et~al.}(2014){De Pontieu}, {Title}, {Lemen}, {Kushner}, {Akin}, {Allard}, {Berger}, {Boerner}, {Cheung}, {Chou}, {Drake}, {Duncan}, {Freeland}, {Heyman}, {Hoffman}, {Hurlburt}, {Lindgren}, {Mathur}, {Rehse}, {Sabolish}, {Seguin}, {Schrijver}, {Tarbell}, {W{\"u}lser}, {Wolfson}, {Yanari}, {Mudge}, {Nguyen-Phuc}, {Timmons}, {van Bezooijen}, {Weingrod}, {Brookner}, {Butcher}, {Dougherty}, {Eder}, {Knagenhjelm}, {Larsen}, {Mansir}, {Phan}, {Boyle}, {Cheimets}, {DeLuca}, {Golub}, {Gates}, {Hertz}, {McKillop}, {Park}, {Perry}, {Podgorski}, {Reeves}, {Saar}, {Testa}, {Tian}, {Weber}, {Dunn}, {Eccles}, {Jaeggli}, {Kankelborg}, {Mashburn}, {Pust}, {Springer}, {Carvalho}, {Kleint}, {Marmie}, {Mazmanian}, {Pereira}, {Sawyer}, {Strong}, {Worden}, {Carlsson}, {Hansteen}, {Leenaarts}, {Wiesmann}, {Aloise}, {Chu}, {Bush}, {Scherrer}, {Brekke}, {Martinez-Sykora}, {Lites}, {McIntosh}, {Uitenbroek}, {Okamoto}, {Gummin}, {Auker}, {Jerram}, {Pool}, \& {Waltham}}]{DePontieu_2014_SoPh}
{De Pontieu}, B., {Title}, A.~M., {Lemen}, J.~R., {et~al.} 2014, \solphys, 289, 2733, \dodoi{10.1007/s11207-014-0485-y}

\bibitem[{{Dere} {et~al.}(1997){Dere}, {Landi}, {Mason}, {Monsignori Fossi}, \& {Young}}]{Dere_1997_AAS}
{Dere}, K.~P., {Landi}, E., {Mason}, H.~E., {Monsignori Fossi}, B.~C., \& {Young}, P.~R. 1997, \aaps, 125, 149, \dodoi{10.1051/aas:1997368}

\bibitem[{{Dey} {et~al.}(2024){Dey}, {Chatterjee}, \& {Erdelyi}}]{Dey_2024_arXiv}
{Dey}, S., {Chatterjee}, P., \& {Erdelyi}, R. 2024, arXiv e-prints, arXiv:2404.16096, \dodoi{10.48550/arXiv.2404.16096}

\bibitem[{{D{\'\i}az Baso} {et~al.}(2021){D{\'\i}az Baso}, {de la Cruz Rodr{\'\i}guez}, \& {Leenaarts}}]{DiazBaso_2021_AA}
{D{\'\i}az Baso}, C.~J., {de la Cruz Rodr{\'\i}guez}, J., \& {Leenaarts}, J. 2021, \aap, 647, A188, \dodoi{10.1051/0004-6361/202040111}

\bibitem[{{Edl{\'e}n}(1943)}]{Edlen_1943_ZAP}
{Edl{\'e}n}, B. 1943, \zap, 22, 30

\bibitem[{{Goodman} \& {Judge}(2012)}]{Goodman_2012_ApJ}
{Goodman}, M.~L., \& {Judge}, P.~G. 2012, \apj, 751, 75, \dodoi{10.1088/0004-637X/751/1/75}

\bibitem[{{Iijima}(2016)}]{Iijima_2016_PhD}
{Iijima}, H. 2016, PhD thesis, \HK{}{Department of Earth and Planetary Environmental Science, The Univ. of Tokyo}

\bibitem[{{Iijima} \& {Yokoyama}(2017)}]{Iijima_2017_ApJ}
{Iijima}, H., \& {Yokoyama}, T. 2017, \apj, 848, 38, \dodoi{10.3847/1538-4357/aa8ad1}

\bibitem[{{Katsukawa} {et~al.}(2020){Katsukawa}, {del Toro Iniesta}, {Solanki}, {Kubo}, {Hara}, {Shimizu}, {Oba}, {Kawabata}, {Tsuzuki}, {Uraguchi}, {Nodomi}, {Shinoda}, {Tamura}, {Suematsu}, {Ishikawa}, {Kano}, {Matsumoto}, {Ichimoto}, {Nagata}, {Quintero Noda}, {Anan}, {Orozco Su{\'a}rez}, {Balaguer Jim{\'e}nez}, {L{\'o}pez Jim{\'e}nez}, {Cobos Carrascosa}, {Feller}, {Riethmueller}, {Gandorfer}, \& {Lagg}}]{Katsukawa_2020_SPIE}
{Katsukawa}, Y., {del Toro Iniesta}, J.~C., {Solanki}, S.~K., {et~al.} 2020, in Society of Photo-Optical Instrumentation Engineers (SPIE) Conference Series, Vol. 11447, Ground-based and Airborne Instrumentation for Astronomy VIII, ed. C.~J. {Evans}, J.~J. {Bryant}, \& K.~{Motohara}, 114470Y, \dodoi{10.1117/12.2561223}

\bibitem[{{Klimchuk}(2006)}]{Klimchuk_2006_SoPh}
{Klimchuk}, J.~A. 2006, \solphys, 234, 41, \dodoi{10.1007/s11207-006-0055-z}

\bibitem[{{Kuniyoshi} {et~al.}(2023){Kuniyoshi}, {Shoda}, {Iijima}, \& {Yokoyama}}]{Kuniyoshi_2023_ApJ}
{Kuniyoshi}, H., {Shoda}, M., {Iijima}, H., \& {Yokoyama}, T. 2023, \apj, 949, 8, \dodoi{10.3847/1538-4357/accbb8}

\bibitem[{{Kuniyoshi} {et~al.}(2024){Kuniyoshi}, {Shoda}, {Morton}, \& {Yokoyama}}]{Kuniyoshi_2024_ApJ}
{Kuniyoshi}, H., {Shoda}, M., {Morton}, R.~J., \& {Yokoyama}, T. 2024, \apj, 960, 118, \dodoi{10.3847/1538-4357/ad1038}

\bibitem[{{Landi} {et~al.}(2012){Landi}, {Del Zanna}, {Young}, {Dere}, \& {Mason}}]{Landi_2012_ApJ}
{Landi}, E., {Del Zanna}, G., {Young}, P.~R., {Dere}, K.~P., \& {Mason}, H.~E. 2012, \apj, 744, 99, \dodoi{10.1088/0004-637X/744/2/99}

\bibitem[{{Leenaarts}(2020)}]{Leenaarts_2020_LRSP}
{Leenaarts}, J. 2020, Living Reviews in Solar Physics, 17, 3, \dodoi{10.1007/s41116-020-0024-x}

\bibitem[{{Lemen} {et~al.}(2012){Lemen}, {Title}, {Akin}, {Boerner}, {Chou}, {Drake}, {Duncan}, {Edwards}, {Friedlaender}, {Heyman}, {Hurlburt}, {Katz}, {Kushner}, {Levay}, {Lindgren}, {Mathur}, {McFeaters}, {Mitchell}, {Rehse}, {Schrijver}, {Springer}, {Stern}, {Tarbell}, {Wuelser}, {Wolfson}, {Yanari}, {Bookbinder}, {Cheimets}, {Caldwell}, {Deluca}, {Gates}, {Golub}, {Park}, {Podgorski}, {Bush}, {Scherrer}, {Gummin}, {Smith}, {Auker}, {Jerram}, {Pool}, {Soufli}, {Windt}, {Beardsley}, {Clapp}, {Lang}, \& {Waltham}}]{Lemen_2012_SoPh}
{Lemen}, J.~R., {Title}, A.~M., {Akin}, D.~J., {et~al.} 2012, \solphys, 275, 17, \dodoi{10.1007/s11207-011-9776-8}

\bibitem[{{Matsumoto} {et~al.}(2023){Matsumoto}, {Kawabata}, {Katsukawa}, {Iijima}, \& {Quintero Noda}}]{Matsumoto_2023_MNRAS}
{Matsumoto}, T., {Kawabata}, Y., {Katsukawa}, Y., {Iijima}, H., \& {Quintero Noda}, C. 2023, \mnras, 523, 974, \dodoi{10.1093/mnras/stad1509}

\bibitem[{{M{\"u}ller} {et~al.}(2020){M{\"u}ller}, {St. Cyr}, {Zouganelis}, {Gilbert}, {Marsden}, {Nieves-Chinchilla}, {Antonucci}, {Auch{\`e}re}, {Berghmans}, {Horbury}, {Howard}, {Krucker}, {Maksimovic}, {Owen}, {Rochus}, {Rodriguez-Pacheco}, {Romoli}, {Solanki}, {Bruno}, {Carlsson}, {Fludra}, {Harra}, {Hassler}, {Livi}, {Louarn}, {Peter}, {Sch{\"u}hle}, {Teriaca}, {del Toro Iniesta}, {Wimmer-Schweingruber}, {Marsch}, {Velli}, {De Groof}, {Walsh}, \& {Williams}}]{Muller_2020_AA}
{M{\"u}ller}, D., {St. Cyr}, O.~C., {Zouganelis}, I., {et~al.} 2020, \aap, 642, A1, \dodoi{10.1051/0004-6361/202038467}

\bibitem[{{Nelson} {et~al.}(2023){Nelson}, {Auch{\`e}re}, {Aznar Cuadrado}, {Barczynski}, {Buchlin}, {Harra}, {Long}, {Parenti}, {Peter}, {Sch{\"u}hle}, {Schwanitz}, {Smith}, {Teriaca}, {Verbeeck}, {Zhukov}, \& {Berghmans}}]{Nelson_2023_AA}
{Nelson}, C.~J., {Auch{\`e}re}, F., {Aznar Cuadrado}, R., {et~al.} 2023, \aap, 676, A64, \dodoi{10.1051/0004-6361/202346144}

\bibitem[{{Panesar} {et~al.}(2021){Panesar}, {Tiwari}, {Berghmans}, {Cheung}, {M{\"u}ller}, {Auchere}, \& {Zhukov}}]{Panesar_2021_ApJ}
{Panesar}, N.~K., {Tiwari}, S.~K., {Berghmans}, D., {et~al.} 2021, \apjl, 921, L20, \dodoi{10.3847/2041-8213/ac3007}

\bibitem[{{Park} {et~al.}(2016){Park}, {Tsiropoula}, {Kontogiannis}, {Tziotziou}, {Scullion}, \& {Doyle}}]{Park_2016_AA}
{Park}, S.~H., {Tsiropoula}, G., {Kontogiannis}, I., {et~al.} 2016, \aap, 586, A25, \dodoi{10.1051/0004-6361/201527440}

\bibitem[{{Parker}(1983)}]{Parker_1983_ApJ}
{Parker}, E.~N. 1983, \apj, 264, 642, \dodoi{10.1086/160637}

\bibitem[{{Pereira} \& {Uitenbroek}(2015)}]{Pereira_2015_AA}
{Pereira}, T. M.~D., \& {Uitenbroek}, H. 2015, \aap, 574, A3, \dodoi{10.1051/0004-6361/201424785}

\bibitem[{{Pesnell} {et~al.}(2012){Pesnell}, {Thompson}, \& {Chamberlin}}]{Pesnell_2012_SoPh}
{Pesnell}, W.~D., {Thompson}, B.~J., \& {Chamberlin}, P.~C. 2012, \solphys, 275, 3, \dodoi{10.1007/s11207-011-9841-3}

\bibitem[{{Pevtsov} {et~al.}(2003){Pevtsov}, {Fisher}, {Acton}, {Longcope}, {Johns-Krull}, {Kankelborg}, \& {Metcalf}}]{Pevtsov_2003_ApJ}
{Pevtsov}, A.~A., {Fisher}, G.~H., {Acton}, L.~W., {et~al.} 2003, \apj, 598, 1387, \dodoi{10.1086/378944}

\bibitem[{{Quintero Noda} {et~al.}(2022){Quintero Noda}, {Schlichenmaier}, {Bellot Rubio}, {L{\"o}fdahl}, {Khomenko}, {Jur{\v{c}}{\'a}k}, {Leenaarts}, {Kuckein}, {Gonz{\'a}lez Manrique}, {Gun{\'a}r}, {Nelson}, {de la Cruz Rodr{\'\i}guez}, {Tziotziou}, {Tsiropoula}, {Aulanier}, {Aboudarham}, {Allegri}, {Alsina Ballester}, {Amans}, {Asensio Ramos}, {Bail{\'e}n}, {Balaguer}, {Baldini}, {Balthasar}, {Barata}, {Barczynski}, {Barreto Cabrera}, {Baur}, {B{\'e}chet}, {Beck}, {Bel{\'\i}o-As{\'\i}n}, {Bello-Gonz{\'a}lez}, {Belluzzi}, {Bentley}, {Berdyugina}, {Berghmans}, {Berlicki}, {Berrilli}, {Berkefeld}, {Bettonvil}, {Bianda}, {Bienes P{\'e}rez}, {Bonaque-Gonz{\'a}lez}, {Braj{\v{s}}a}, {Bommier}, {Bourdin}, {Burgos Mart{\'\i}n}, {Calchetti}, {Calcines}, {Calvo Tovar}, {Campbell}, {Carballo-Mart{\'\i}n}, {Carbone}, {Carlin}, {Carlsson}, {Castro L{\'o}pez}, {Cavaller}, {Cavallini}, {Cauzzi}, {Cecconi}, {Chulani}, {Cirami}, {Consolini}, {Coretti}, {Cosentino}, {C{\'o}zar-Castellano}, {Dalmasse}, {Danilovic}, {De Juan
  Ovelar}, {Del Moro}, {del Pino Alem{\'a}n}, {del Toro Iniesta}, {Denker}, {Dhara}, {Di Marcantonio}, {D{\'\i}az Baso}, {Diercke}, {Dineva}, {D{\'\i}az-Garc{\'\i}a}, {Doerr}, {Doyle}, {Erdelyi}, {Ermolli}, {Escobar Rodr{\'\i}guez}, {Esteban Pozuelo}, {Faurobert}, {Felipe}, {Feller}, {Feijoo Amoedo}, {Femen{\'\i}a Castell{\'a}}, {Fernandes}, {Ferro Rodr{\'\i}guez}, {Figueroa}, {Fletcher}, {Franco Ordovas}, {Gafeira}, {Gardenghi}, {Gelly}, {Giorgi}, {Gisler}, {Giovannelli}, {Gonz{\'a}lez}, {Gonz{\'a}lez}, {Gonz{\'a}lez-Cava}, {Gonz{\'a}lez Garc{\'\i}a}, {G{\"o}m{\"o}ry}, {Gracia}, {Grauf}, {Greco}, {Grivel}, {Guerreiro}, {Guglielmino}, {Hammerschlag}, {Hanslmeier}, {Hansteen}, {Heinzel}, {Hern{\'a}ndez-Delgado}, {Hern{\'a}ndez Su{\'a}rez}, {Hidalgo}, {Hill}, {Hizberger}, {Hofmeister}, {J{\"a}gers}, {Janett}, {Jarolim}, {Jess}, {Jim{\'e}nez Mej{\'\i}as}, {Jolissaint}, {Kamlah}, {Kapit{\'a}n}, {Ka{\v{s}}parov{\'a}}, {Keller}, {Kentischer}, {Kiselman}, {Kleint}, {Klvana}, {Kontogiannis}, {Krishnappa},
  {Ku{\v{c}}era}, {Labrosse}, {Lagg}, {Landi Degl'Innocenti}, {Langlois}, {Lafon}, {Laforgue}, {Le Men}, {Lepori}, {Lepreti}, {Lindberg}, {Lilje}, {L{\'o}pez Ariste}, {L{\'o}pez Fern{\'a}ndez}, {L{\'o}pez Jim{\'e}nez}, {L{\'o}pez L{\'o}pez}, {Manso Sainz}, {Marassi}, {Marco de la Rosa}, {Marino}, {Marrero}, {Mart{\'\i}n}, {Mart{\'\i}n G{\'a}lvez}, {Mart{\'\i}n Hernando}, {Masciadri}, {Mart{\'\i}nez Gonz{\'a}lez}, {Matta-G{\'o}mez}, {Mato}, {Mathioudakis}, {Matthews}, {Mein}, {Merlos Garc{\'\i}a}, {Moity}, {Montilla}, {Molinaro}, {Molodij}, {Montoya}, {Munari}, {Murabito}, {N{\'u}{\~n}ez Cagigal}, {Oliviero}, {Orozco Su{\'a}rez}, {Ortiz}, {Padilla-Hern{\'a}ndez}, {Pa{\'e}z Ma{\~n}{\'a}}, {Paletou}, {Pancorbo}, {Pastor Ca{\~n}edo}, {Pastor Yabar}, {Peat}, {Pedichini}, {Peixinho}, {Pe{\~n}ate}, {P{\'e}rez de Taoro}, {Peter}, {Petrovay}, {Piazzesi}, {Pietropaolo}, {Pleier}, {Poedts}, {P{\"o}tzi}, {Podladchikova}, {Prieto}, {Quintero Nehrkorn}, {Ramelli}, {Ramos Sapena}, {Rasilla}, {Reardon}, {Rebolo}, {Regalado
  Olivares}, {Reyes Garc{\'\i}a-Talavera}, {Riethm{\"u}ller}, {Rimmele}, {Rodr{\'\i}guez Delgado}, {Rodr{\'\i}guez Gonz{\'a}lez}, {Rodr{\'\i}guez-Losada}, {Rodr{\'\i}guez Ramos}, {Romano}, {Roth}, {Rouppe van der Voort}, {Rudawy}, {Ruiz de Galarreta}, {Ryb{\'a}k}, {Salvade}, {S{\'a}nchez-Capuchino}, {S{\'a}nchez Rodr{\'\i}guez}, {Sangiorgi}, {Say{\`e}de}, {Scharmer}, {Scheiffelen}, {Schmidt}, {Schmieder}, {Scir{\`e}}, {Scuderi}, {Siegel}, {Sigwarth}, {Sim{\~o}es}, {Snik}, {Sliepen}, {Sobotka}, {Socas-Navarro}, {Sola La Serna}, {Solanki}, {Soler Trujillo}, {Soltau}, {Sordini}, {Sosa M{\'e}ndez}, {Stangalini}, {Steiner}, {Stenflo}, {{\v{S}}t{\v{e}}p{\'a}n}, {Strassmeier}, {Sudar}, {Suematsu}, {S{\"u}tterlin}, {Tallon}, {Temmer}, {Tenegi}, {Tritschler}, {Trujillo Bueno}, {Turchi}, {Utz}, {van Harten}, {van Noort}, {van Werkhoven}, {Vansintjan}, {Vaz Cedillo}, {Vega Reyes}, {Verma}, {Veronig}, {Viavattene}, {Vitas}, {V{\"o}gler}, {von der L{\"u}he}, {Volkmer}, {Waldmann}, {Walton}, {Wisniewska}, {Zeman},
  {Zeuner}, {Zhang}, {Zuccarello}, \& {Collados}}]{Quintero_2022_AA}
{Quintero Noda}, C., {Schlichenmaier}, R., {Bellot Rubio}, L.~R., {et~al.} 2022, \aap, 666, A21, \dodoi{10.1051/0004-6361/202243867}

\bibitem[{{Rempel}(2017)}]{Rempel_2017_ApJ}
{Rempel}, M. 2017, \apj, 834, 10, \dodoi{10.3847/1538-4357/834/1/10}

\bibitem[{{Rimmele} {et~al.}(2020){Rimmele}, {Warner}, {Keil}, {Goode}, {Kn{\"o}lker}, {Kuhn}, {Rosner}, {McMullin}, {Casini}, {Lin}, {W{\"o}ger}, {von der L{\"u}he}, {Tritschler}, {Davey}, {de Wijn}, {Elmore}, {Fehlmann}, {Harrington}, {Jaeggli}, {Rast}, {Schad}, {Schmidt}, {Mathioudakis}, {Mickey}, {Anan}, {Beck}, {Marshall}, {Jeffers}, {Oschmann}, {Beard}, {Berst}, {Cowan}, {Craig}, {Cross}, {Cummings}, {Donnelly}, {de Vanssay}, {Eigenbrot}, {Ferayorni}, {Foster}, {Galapon}, {Gedrites}, {Gonzales}, {Goodrich}, {Gregory}, {Guzman}, {Guzzo}, {Hegwer}, {Hubbard}, {Hubbard}, {Johansson}, {Johnson}, {Liang}, {Liang}, {McQuillen}, {Mayer}, {Newman}, {Onodera}, {Phelps}, {Puentes}, {Richards}, {Rimmele}, {Sekulic}, {Shimko}, {Simison}, {Smith}, {Starman}, {Sueoka}, {Summers}, {Szabo}, {Szabo}, {Wampler}, {Williams}, \& {White}}]{Rimmele_2020_SoPh}
{Rimmele}, T.~R., {Warner}, M., {Keil}, S.~L., {et~al.} 2020, \solphys, 295, 172, \dodoi{10.1007/s11207-020-01736-7}

\bibitem[{{Rochus} {et~al.}(2020){Rochus}, {Auch{\`e}re}, {Berghmans}, {Harra}, {Schmutz}, {Sch{\"u}hle}, {Addison}, {Appourchaux}, {Aznar Cuadrado}, {Baker}, {Barbay}, {Bates}, {BenMoussa}, {Bergmann}, {Beurthe}, {Borgo}, {Bonte}, {Bouzit}, {Bradley}, {B{\"u}chel}, {Buchlin}, {B{\"u}chner}, {Cab{\'e}}, {Cadiergues}, {Chaigneau}, {Chares}, {Choque Cortez}, {Coker}, {Condamin}, {Coumar}, {Curdt}, {Cutler}, {Davies}, {Davison}, {Defise}, {Del Zanna}, {Delmotte}, {Delouille}, {Dolla}, {Dumesnil}, {D{\"u}rig}, {Enge}, {Fran{\c{c}}ois}, {Fourmond}, {Gillis}, {Giordanengo}, {Gissot}, {Green}, {Guerreiro}, {Guilbaud}, {Gyo}, {Haberreiter}, {Hafiz}, {Hailey}, {Halain}, {Hansotte}, {Hecquet}, {Heerlein}, {Hellin}, {Hemsley}, {Hermans}, {Hervier}, {Hochedez}, {Houbrechts}, {Ihsan}, {Jacques}, {J{\'e}r{\^o}me}, {Jones}, {Kahle}, {Kennedy}, {Klaproth}, {Kolleck}, {Koller}, {Kotsialos}, {Kraaikamp}, {Langer}, {Lawrenson}, {Le Clech'}, {Lenaerts}, {Liebecq}, {Linder}, {Long}, {Mampaey}, {Markiewicz-Innes}, {Marquet},
  {Marsch}, {Matthews}, {Mazy}, {Mazzoli}, {Meining}, {Meltchakov}, {Mercier}, {Meyer}, {Monecke}, {Monfort}, {Morinaud}, {Moron}, {Mountney}, {M{\"u}ller}, {Nicula}, {Parenti}, {Peter}, {Pfiffner}, {Philippon}, {Phillips}, {Plesseria}, {Pylyser}, {Rabecki}, {Ravet-Krill}, {Rebellato}, {Renotte}, {Rodriguez}, {Roose}, {Rosin}, {Rossi}, {Roth}, {Rouesnel}, {Roulliay}, {Rousseau}, {Ruane}, {Scanlan}, {Schlatter}, {Seaton}, {Silliman}, {Smit}, {Smith}, {Solanki}, {Spescha}, {Spencer}, {Stegen}, {Stockman}, {Szwec}, {Tamiatto}, {Tandy}, {Teriaca}, {Theobald}, {Tychon}, {van Driel-Gesztelyi}, {Verbeeck}, {Vial}, {Werner}, {West}, {Westwood}, {Wiegelmann}, {Willis}, {Winter}, {Zerr}, {Zhang}, \& {Zhukov}}]{Rochus_2020_AA}
{Rochus}, P., {Auch{\`e}re}, F., {Berghmans}, D., {et~al.} 2020, \aap, 642, A8, \dodoi{10.1051/0004-6361/201936663}

\bibitem[{{Scharmer} {et~al.}(2003){Scharmer}, {Bjelksjo}, {Korhonen}, {Lindberg}, \& {Petterson}}]{Scharmer_2003_SPIE}
{Scharmer}, G.~B., {Bjelksjo}, K., {Korhonen}, T.~K., {Lindberg}, B., \& {Petterson}, B. 2003, in Society of Photo-Optical Instrumentation Engineers (SPIE) Conference Series, Vol. 4853, Innovative Telescopes and Instrumentation for Solar Astrophysics, ed. S.~L. {Keil} \& S.~V. {Avakyan}, 341--350, \dodoi{10.1117/12.460377}

\bibitem[{{Scharmer} {et~al.}(2008){Scharmer}, {Narayan}, {Hillberg}, {de la Cruz Rodriguez}, {L{\"o}fdahl}, {Kiselman}, {S{\"u}tterlin}, {van Noort}, \& {Lagg}}]{Scharmer_2008_ApJ}
{Scharmer}, G.~B., {Narayan}, G., {Hillberg}, T., {et~al.} 2008, \apjl, 689, L69, \dodoi{10.1086/595744}

\bibitem[{{Shetye} {et~al.}(2019){Shetye}, {Verwichte}, {Stangalini}, {Judge}, {Doyle}, {Arber}, {Scullion}, \& {Wedemeyer}}]{Shetye_2019_ApJ}
{Shetye}, J., {Verwichte}, E., {Stangalini}, M., {et~al.} 2019, \apj, 881, 83, \dodoi{10.3847/1538-4357/ab2bf9}

\bibitem[{{Silva} {et~al.}(2024){Silva}, {Verth}, {Rempel}, {Ballai}, {Jafarzadeh}, \& {Fedun}}]{Silva_2024_ApJ}
{Silva}, S. S.~A., {Verth}, G., {Rempel}, E.~L., {et~al.} 2024, \apj, 963, 10, \dodoi{10.3847/1538-4357/ad1403}

\bibitem[{{Sukhorukov} \& {Leenaarts}(2017)}]{2017A&A...597A..46S}
{Sukhorukov}, A.~V., \& {Leenaarts}, J. 2017, \aap, 597, A46, \dodoi{10.1051/0004-6361/201629086}

\bibitem[{{Tziotziou} {et~al.}(2018){Tziotziou}, {Tsiropoula}, {Kontogiannis}, {Scullion}, \& {Doyle}}]{Tziotziou_2018_AA}
{Tziotziou}, K., {Tsiropoula}, G., {Kontogiannis}, I., {Scullion}, E., \& {Doyle}, J.~G. 2018, \aap, 618, A51, \dodoi{10.1051/0004-6361/201833101}

\bibitem[{{Uitenbroek}(2001)}]{2001ApJ...557..389U}
{Uitenbroek}, H. 2001, \apj, 557, 389, \dodoi{10.1086/321659}

\bibitem[{{Van Doorsselaere} {et~al.}(2016){Van Doorsselaere}, {Antolin}, {Yuan}, {Reznikova}, \& {Magyar}}]{VanDoorsselaere_2016_FrASS}
{Van Doorsselaere}, T., {Antolin}, P., {Yuan}, D., {Reznikova}, V., \& {Magyar}, N. 2016, Frontiers in Astronomy and Space Sciences, 3, 4, \dodoi{10.3389/fspas.2016.00004}

\bibitem[{{Van Doorsselaere} {et~al.}(2020){Van Doorsselaere}, {Srivastava}, {Antolin}, {Magyar}, {Vasheghani Farahani}, {Tian}, {Kolotkov}, {Ofman}, {Guo}, {Arregui}, {De Moortel}, \& {Pascoe}}]{VanDoorsselaere_2020_SSRv}
{Van Doorsselaere}, T., {Srivastava}, A.~K., {Antolin}, P., {et~al.} 2020, \ssr, 216, 140, \dodoi{10.1007/s11214-020-00770-y}

\bibitem[{{Wedemeyer-B{\"o}hm} {et~al.}(2012){Wedemeyer-B{\"o}hm}, {Scullion}, {Steiner}, {Rouppe van der Voort}, {de La Cruz Rodriguez}, {Fedun}, \& {Erd{\'e}lyi}}]{Wedemeyer_2012_Nature}
{Wedemeyer-B{\"o}hm}, S., {Scullion}, E., {Steiner}, O., {et~al.} 2012, \nat, 486, 505, \dodoi{10.1038/nature11202}

\bibitem[{{Withbroe} \& {Noyes}(1977)}]{Withbroe_1977_ARAA}
{Withbroe}, G.~L., \& {Noyes}, R.~W. 1977, \araa, 15, 363, \dodoi{10.1146/annurev.aa.15.090177.002051}

\end{thebibliography}
\end{document}